\documentclass{aa}

\usepackage[tbtags,fleqn]{amsmath}
\usepackage{amsfonts}
\usepackage{graphicx}
\usepackage{pstricks, multido, pst-plot}

\usepackage{natbib}
\bibpunct{(}{)}{;}{a}{}{,}

\DeclareMathOperator{\sgn}{sgn}
\DeclareMathOperator{\Var}{Var}
\DeclareMathOperator{\Cov}{Cov}

\begin{document}

\title{Mapping the interstellar dust with near-infrared
  observations: An optimized multi-band technique}
\author{Marco Lombardi\inst{1} and Jo\~ao Alves\inst{2}}
\authorrunning{M. Lombardi \& J. Alves}
\titlerunning{Mapping the interstellar dust with NIR observations}
\offprints{M. Lombardi}
\mail{lombardi@astro.uni-bonn.de}
\institute{%
  Instit\"ut f\"ur Astrophysik und Extraterrestrische Forschung,
  Auf dem H\"ugel 71, D-53121 Bonn, Germany
  \and
  European Southern Observatory,
  Karl-Schwarzschild-Stra\ss e 2,
  D-85748 Garching bei M\"unchen, Germany}
\date{Received: 29 May 2001 / Accepted: 30 July 2001}

\abstract{%
  We generalize the technique of \citet{1994ApJ...429..694L} to map
  dust column density through a molecular cloud (\textsc{Nice}) to an
  optimized multi-band technique (\textsc{Nicer}) that can be applied
  to any multi-band survey of molecular clouds.  We present a first
  application to a $\sim 625 \mbox{ deg}^2$ subset of the Two Micron
  All Sky Survey (2MASS) data and show that when compared to
  \textsc{Nice}, the optimized \textsc{Nicer} technique (i) achieves
  the same extinction peak values, (ii) improves the noise variance of
  the map by a factor of 2 and (iii) is able to reach $3 \, \sigma$
  dust extinction measurements as low as $A_V \simeq 0.5 \mbox{
    magnitudes}$, better than or equivalent to classical optical star
  count techniques and below the threshold column density for the
  formation of CO, the brightest $\mbox{H}_2$ tracer in
  radio-spectroscopy techniques.  The application of the
  \textsc{Nicer} techniques to near-infrared data obtained with a 8
  meter-class telescope with a state-of-the-art NIR camera, such as
  the VLT-ISAAC combination, will be able to achieve dynamic ranges
  from below $10^{21} \mbox{ protons cm}^{-2}$ to over $10^{23} \mbox{
    protons cm}^{-2}$ ($A_V$ in the range $[0.3, 60]$) and spatial
  resolutions $< 10''$, making use of a single and straightforward
  column density tracer, extinction by interstellar dust.
  } \maketitle

%

\section{Introduction}
\label{sec:introduction}

Dark clouds remain one of the least understood objects in the
Universe.  Although they are ubiquitous in galaxies like the Milky
Way, we still do not have a basic understanding of how they relate to
the more diffuse interstellar medium (ISM), of how some of them form
stars and planets, or how they vanish.  We do know today that they are
the coldest objects known ($T \simeq 10 \mbox{ K}$), which
unfortunately implies that they are also one of the hardest to observe
and study.
 
Before the discovery of emission from molecules in dark clouds
(\citealp{1970ApJ...161L..43W}, after which these clouds became known
as molecular clouds), the general technique used to study the density
structure of these objects relied on statistical analysis of numbers
of dust extincted stars in the background of a cloud.  This approach,
introduced by \citet{1923AN....219..109W}, is known as the star count
method and was later refined and intensively applied by
\citet{1937dss..book.....B, 1956AJ.....61..309B}.  In this technique,
the main mass component of a molecular cloud, $\mathrm{H}_2$, is
traced by the dust in the cloud, an assumption later proved to be
valid by observations.  The main feat of the star count method is its
straightforward essence and its sensitivity to low column density
regions with $A_V$ in the range $[0.5, 3]$ magnitudes, i.e., from
column densities between $\sim 1 \cdot 10^{21}$ and $\sim 8 \cdot
10^{21} \mbox{ protons cm}^{-2}$ (\citealp{1978AJ.....83..241R,
  1978AJ.....83..363D, 1986A&A...160..157M, 1988A&AS...76..347G}; see
\citealp{1999ApJ...517..264A} for a multi-technique approach).

Almost everything we know today about molecular cloud structure has
been derived from radio observations of molecular tracers (e.g., CO,
CS, $\mathrm{NH}_3$) of the undetectable $\mathrm{H}_2$
\citep{1996IAUS..170..387L, 1999osps.conf....3B, 1999osps.conf...67M}
and in more recent years through dust thermal emission that can be
detected by radio continuum techniques \citep{2000prpl.conf...59A,
  2000ApJ...545..327J}.  Radio spectroscopy and continuum techniques
are able to probe deeper into molecular clouds, well past column
densities of $10^{22} \mbox{ protons cm}^{-2}$, an order of magnitude
improvement over classical star count techniques.  Moreover,
radio-spectroscopy offers an unique view of the dynamical structure of
molecular clouds \citep[e.g.,][]{1999ApJ...524..887R}, something that
none of the absorption or emission continuum techniques can,
evidently, assess.  However, the interpretation of results derived
from radio observations is not always straightforward
\citep[e.g.][]{1999ApJ...515..265A, 2000ApJ...530..851C}.  Several
poorly constrained effects inherent to these techniques (e.g.,
deviations from local thermodynamic equilibrium, opacity variations,
chemical evolution, small-scale structure, depletion of molecules,
unknown emissivity properties of the dust, unknown dust temperature)
make the construction of an unambiguous picture of the physical
structure of these objects a very difficult task.  The price of being
able to probe deeper into a molecular cloud with radio line and
continuum techniques is a complicated data analysis, not free from
ambiguities.

It has long been recognized that infrared color excess can be used to
obtain a reliable estimate of the extinction through a molecular cloud
\citep[see, e.g.,][]{1977PASP...89..597B, 1980ApJ...242..132J,
  1982ApJ...262..590F, 1984ApJ...282..675J, 1986MNRAS.223..341C}.  The
deployment of sensitive, large format infrared array cameras on large
telescopes, however, has allowed the direct measurement of
line-of-sight dust extinction toward thousands of individual
background stars observed through a molecular cloud.  Such
measurements are free from the complications that plague
molecular-line or dust emission data and enable detailed maps of cloud
density structure to be constructed. In all fairness, all density
tracer methods rely on the assumption of an universal gas-to-dust
ratio, a fair assumption on theoretical grounds that still remains to
be demonstrated for cold dense molecular clouds.
\citet{1994ApJ...429..694L} pioneered a technique, the Near-Infrared
Color Excess (\textsc{Nice}) technique, for measuring and mapping the
distribution of dust through a molecular cloud using data obtained in
large-scale, multi-wavelength, infrared imaging surveys.  This method
combines measurements of near-infrared color excess to directly
measure extinctions (as opposed to the statistical measurement in the
star count method) and map the dust column density distribution
through a cloud.  Moreover, the measurements can be made at
significantly higher angular resolutions and substantially greater
optical depths than previously thought possible.  The efficacy of this
technique was demonstrated with the study of the dark cloud L~977
\citep{1998ApJ...506..292A}, IC~5146 \citep{1999ApJ...512..250L}, and
Barnard~68 \citep{2001Natur.409..159A} where through straightforward
analysis of nearly $7000$ infrared sources background to these clouds
produced detailed maps of the extinction to optical depths and spatial
resolution an order of magnitude higher than previously possible
($A_V$ in the range $[1, 40]$ magnitudes, spatial resolution down to
$10 \mbox{ arcsec}$).

\subsection{The large view on molecular clouds}

Nearby galactic molecular cloud complexes represent our best chance to
understand molecular cloud structure.  These complexes have relatively
large sizes (from a few to tens of parsec) and are hence fairly
extended, typically stretching several to tens of degrees across the
sky.  A large view of entire molecular cloud complexes, with
sufficient resolution and dynamic range, is the only way to put these
clouds in their Galactic context and address the questions on their
origin and fate.  Census of entire molecular cloud complexes are rare
(CO line emission: \citealp{2001ApJ...547..792D, 1999PASJ...51..745F};
IRAS opacity maps: \citealp{1995Ap&SS.224..199W}; optical star counts:
\citealp{1999A&A...345..965C}), have typical dynamical ranges of
$10^{21} \mbox{ protons cm}^{-2}$ to below or about $10^{22} \mbox{
  protons cm}^{-2}$, but suffer from the caveats inherent to the
respective used techniques (see above). There is a clear need for a
new large scale mapping technique of molecular cloud complexes, that
makes use of a more reliable column density tracer, and is able to
offer an extended dynamical range.

In this paper we generalize Lada's \textsc{Nice} technique to an
optimized multi-band technique, the Near-Infrared Color Excess
Revisited technique (\textsc{Nicer}).  Although inspired by the Two
Micron All Sky Survey (2MASS) \citep{1994ExA.....3...65K}, and by the
possibility of all sky dust extinction mapping, this technique can be
applied to any multi-band survey of molecular clouds [e.g., DENIS
\citep{1997Msngr..87...27E}, SDSS \citep{2000AJ....120.1579Y}, or any
combination of multi-wavelength catalogs].  We present a first
application to 2MASS data, and show that \textsc{Nicer} improves the
noise variance of a map by a factor of 2 when compared to
\textsc{Nice}, and is able to reach $3 \, \sigma$ dust extinction
measurements of $A_V \simeq 0.5$ magnitudes, better than or equivalent
to optical star count techniques and below the threshold column
density for the formation of CO under typical temperatures and
densities \citep{1988ApJ...334..771V, 1992IAUS..150..143V}. This
latter point is highly relevant as CO is the brightest $\mbox{H}_2$
tracer in radio-spectroscopy techniques.  The application of
\textsc{Nicer} to near-infrared (NIR) data obtained with an 8
meter-class telescope outfitted with a state-of-the-art NIR camera,
e.g.\ the VLT-ISAAC combination, will be able to achieve column
density dynamic ranges from below $10^{21} \mbox{ protons cm}^{-2}$ to
over $10^{23} \mbox{ protons cm}^{-2}$ ($A_V$ in the range $[0.3,60]$
magnitudes), on spatial resolutions $< 10''$, making use of a single
and straightforward column density tracer, extinction by dust.

The paper is organized as follows: In Sect.~2 we review the
\textsc{Nice} technique and discuss its limitations, in Sect.~3 we
give a detailed presentation of \textsc{Nicer}, in Sect.~4 we apply
both techniques to the same 2MASS data subset, and present the
conclusions in Sect.~5.

\section{Near-infrared color excess}
\label{sec:near-infrared-color}

The difference between the observed and the intrinsic color of a
background star provides information on the optical depth of the
cloud.  For example, for the near-infrared and assuming a
\textit{normal reddening law\/} \citep{1985ApJ...288..618R}, we have
\begin{equation}
  \label{eq:1}
  (H - K)^\mathrm{obs} = (H - K)^\mathrm{tr} + 0.063 \cdot A_V \; ,
\end{equation}

where H and K represent the H and K-band (centered at 1.65$\mu$m and
2.2$\mu$m, respectively) the superscripts ``obs'' and ``tr'' denote
observed and true (intrinsic) quantities, respectively.  The visual
extinction $A_V$ is often used as a measure of dust column density and
will be the main target of our investigations.  This quantity is
relevant because, as suggested by several studies
\citep{1955ApJ...121..559L, 1974ApJ...187..243J, 1978ApJ...224..132B},
it is directly related to the projected mass density of the cloud.  In
particular, it has been shown that most clouds have the same
gas-to-dust ratio, so that $N(H_2) = A_V \cdot 1.9\cdot 10^{21} \mbox{
  protons cm}^{-2} \mbox{ mag}^{-1}$ or, equivalently for the
projected two-dimensional density, $\kappa = A_V \cdot 15 \mbox{
  M}_\odot \mbox{ pc}^{-2}$.

Equation~\eqref{eq:1} would not be particularly useful if the
intrinsic star colors (which are generally unknown) would span a large
range of values.  In reality, infrared colors of stars have a
relatively small scatter (e.g., $\sigma (H-K)^\mathrm{tr} \simeq
0.08$), so that it is sensible to take the average of
Eq.~\eqref{eq:1}, obtaining
\begin{equation}
  \label{eq:2}
  A_V = 15.87 \cdot \Bigl[ \bigl\langle (H - K)^\mathrm{obs}
  \bigr\rangle -  \bigl\langle (H - K)^\mathrm{tr} \bigr\rangle
  \Bigr] \; .
\end{equation}
It is not difficult to estimate $\bigl\langle (H - K)^\mathrm{tr}
\bigr\rangle$ using, for example, a control field where the extinction
is negligible.  Hence, in principle, we can use Eq.~\eqref{eq:2} with
$\bigl\langle (H - K)^\mathrm{obs} \bigr\rangle$ replaced by the
observed color of each \textit{background\/} star and obtain an
\textit{unbiased\/} estimate for the cloud column density in the
direction of the star.  In practice, this method suffers from two main
limitations: (i) The use of the observed color of each star in lieu of
$\bigl\langle (H - K)^\mathrm{obs} \bigr\rangle$ introduces a source
of noise in the estimate of $A_V$; (ii) It may \textit{not be
  easy\/} to establish if a given star is a background object, i.e.\ 
is farther than the cloud.  Traditionally, both limitations are dealt
with by using two simple techniques.  The noise on the measured column
density, which is ultimately introduced by the intrinsic scatter on
the colors of stars, is reduced by averaging angularly close objects
(which, hopefully, are subject to the same amount of extinction; see
however below).  Thus, normally the extinction is estimated using the
equation
\begin{equation}
  \label{eq:3}
  \hat A_V = 15.87 \biggl[ \frac{1}{N} \sum_{n=1}^N (H -
  K)^\mathrm{obs}_i - \bigl\langle (H - K)^\mathrm{tr} \bigr\rangle
  \biggr] \; ,
\end{equation}
where the sum is performed on a set of $N$ angularly close stars.
This is the approach taken by the \textsc{Nice} method, originally
described by \citet{1994ApJ...429..694L}.  Regarding the second
problem, foreground stars can be recognized from their color, which is
significantly bluer than angularly nearby stars (often the number of
foreground stars is much smaller than the number of background stars).

It is not difficult to recognize that the two problems discussed above
are intimately related.  Suppose, for example, that we are
investigating a cloud with low column density.  Then, because of the
intrinsic scatter of star colors and of photometric errors, we could
take a foreground star as a background object just because its is
observed redder than expected.  The situation can be even more
difficult for embedded stars.  Clearly, a \textit{perfect\/}
discrimination between foreground and background objects is feasible
only in the unrealistic case where the column density can be measured
\textit{without any error}.  This is not possible, but at least we can
try to take care of both problems in a uniform way.  This is the point
of view adopted by our optimized method for extinction estimate.

\section{The optimized method}
\label{sec:optimized-method}

\begin{figure}[t!]
 \begin{center}
   \ {}
   \vskip 5truecm
   \ {} 
   \caption{$JHK$ color-color of one hundred stars (sketch).  The left plot
     shows the colors of stars which are not subject to extinction.
     The ellipse encloses a $3\sigma$ confidence region; note that the
     ellipse is almost exactly vertical oriented, i.e.\ $J-H$
     correlates only weakly with $H-K$.  The right plot shows the same
     stars as observed through a cloud with $A_V = 5$ magnitudes (a
     normal reddening law is assumed).}
   \label{fig:1}
 \end{center}
\end{figure}

In this section we present \textsc{Nicer}, the Near-Infrared Color
Excess method Revisited, an optimized, multi-band method for
extinction estimation based on the near infrared excess of background
starlight.  The method has been designed by considering separately the two
steps needed to make an extinction map:
\begin{itemize}
\item Local extinction estimate for each star;
\item Spatial smoothing of individual stars estimates.
\end{itemize}
We describe these two steps in Sects.~\ref{sec:local-absorption} and
\ref{sec:spatial-smoothing}.

\subsection{Local extinction}
\label{sec:local-absorption}

As discussed in Sect.~\ref{sec:near-infrared-color}, the cloud column
density $A_V$ is normally measured by comparing the $H - K$ color for
stars observed through the cloud with the same color for stars for
which no reddening is expected [see Eq.~\eqref{eq:3}].  Clearly, other
possibilities are also viable.  For instance, we could choose the $J -
H$ color and use the expression
\begin{equation}
  \label{eq:4}
  \hat A_V = 9.35 \biggl[ \frac{1}{N} \sum_{n=1}^N (J -
  H)^\mathrm{obs}_i - \bigl\langle (J - H)^\mathrm{tr} \bigr\rangle
  \biggr] \; .
\end{equation}
Should we use Eq.~\eqref{eq:3} or \eqref{eq:4} to infer the cloud
column density?  In both cases, we expect some error on the estimate
of $A_V$.  The final error on $A_V$ depends critically on
three factors:
\begin{enumerate}
\item The scatter of the intrinsic colors: Normally, stars present a
  larger scatter in $J - H$ than in $H - K$ (see Fig.~\ref{fig:1}).
\item The photometric error on the individual bands: Typically errors
  on $J$ are smaller than errors on $K$ (except along lines-of-sight
  through dense clouds, where extinction makes stars fainter in
  $J$).
\item The numerical coefficient used to convert color excess into
  $A_V$ extinction:  This coefficient is larger for Eq.~\eqref{eq:3}
  ($15.87$ vs.\ $9.35$).
\end{enumerate}
Hence, on the one hand we are encouraged to use Eq.~\eqref{eq:3}
because of the smaller scatter in colors; on the other hand we want to
avoid this estimator because of the larger numerical coefficient and
larger photometric errors.  In conclusion, this Heuristic discussion
does not lead to a definitive answer regarding the choice between the
two estimators \eqref{eq:3} and \eqref{eq:4}.  This suggests that a
more quantitative approach is needed to study the problem and decide
which estimator should be used.  Actually, the analysis we are going
to carry out will provide a much more interesting result and will show
that a clever use of all available infrared bands produces
significantly more accurate results than the individual estimators
\eqref{eq:3} or \eqref{eq:4}.  [In the following we will consider the
typical case of three bands $J$, $H$, and $K$ available; The
generalization to more bands is trivial.]

\begin{figure}[t!]
 \begin{center}
   \ {}
   \vskip 5truecm
   \ {} 
   \caption{For each star, \textsc{Nicer} obtains an estimate of the
     extinction $A_V$ using a maximum-likelihood technique.  The star
     colors and the relative covariance (represented in the figure in
     the left panel by the small ellipse to the top right) are
     obtained from the observations.  Note that correlation of colors
     is expected [see Eq.~\eqref{eq:10}].  The net effect of
     photometric errors is a widening (solid ellipse) of the intrinsic
     color scatter (dashed ellipse) [Eq.~\eqref{eq:11} and following
     discussion].  To estimate the extinction to the star
     \textsc{Nicer} than finds the ellipse center along the reddening
     line (see figure to the right) that is closer to the observed
     colors of the star [Eq.~\eqref{eq:12}].}
   \label{fig:2}
 \end{center}
\end{figure}

With the three bands at our disposal we can make two independent
colors, $c_1 = J - H$ and $c_2 = H - K$.  For each $c_i$ (with $i = 1,
2$), we can write the relationship between the observed color
$c_i^\mathrm{obs}$ and the true one $c_i^\mathrm{tr}$ (i.e., the color
that would be observed if no extinction were present) as
\begin{equation}
  \label{eq:5}
  c_i^\mathrm{obs} = c_i^\mathrm{tr} + k_i A_V + \epsilon_i \; .
\end{equation}
Here $k_i = E_i/A_V$ is the ratio between the color excess on the band
$i$ and the extinction on the $V$ band; for the colors considered here
we have $k_1 = 1/9.35$ and $k_2 = 1/15.87$.  The extra term
$\epsilon_i$ above represents the noise on the colors, i.e.\ the
result of photometric error on the estimate of $J-H$ and $H-K$.  Let
us restrict the discussion to estimators of $A_V$ which are linear in
the observed colors $c_i^\mathrm{obs}$, i.e.\ of the form
\begin{equation}
  \label{eq:6}
  \hat A_V = a + b_1 c_1^\mathrm{obs} + b_2 c_2^\mathrm{obs} \; .
\end{equation}
The coefficients $a$, $b_1$, and $b_2$ need to be determined so as to
satisfy two conditions (see Fig.~\ref{fig:2}):
\begin{enumerate}
\item The estimator is \textit{unbiased}, i.e.\ its expected value is
  the true extinction $A_V$.
\item The estimator has \textit{minimum variance}.
\end{enumerate}
The first condition is equivalent to the equations
\begin{align}
  \label{eq:7}
  b_1 k_1 + b_2 k_2 &= 1 \; , & a + b_1 \langle c_1^\mathrm{tr}
  \rangle + b_2 \langle c_2^\mathrm{tr} \rangle &= 0 \; .
\end{align}
The variance of the estimator $\hat A_V$ can be evaluated from the
expression
\begin{equation}
  \label{eq:8}
  \hat A_V - \bigl\langle \hat A_V \bigr\rangle = \sum_i b_i \bigl(
  c_i^\mathrm{tr} - \langle c_i^\mathrm{tr} \rangle \bigr) + \sum_i
  b_i \epsilon_i \; .
\end{equation}
As a result, we can immediately write
\begin{equation}
  \label{eq:9}
  \Var \bigl( \hat A_V \bigr) = \sum_{i,j} b_i b_j
  \Cov_{ij}(c^\mathrm{tr}) + \sum_{i,j} b_i b_j \Cov_{ij}(\epsilon) \; .
\end{equation}
The covariance matrices introduced in this equation represent two
different sources of noise.  The first matrix,
$\Cov_{ij}(c^\mathrm{tr})$ is the intrinsic scatter of star colors,
which clearly makes the determination of the extinction $A_V$
uncertain (for example, if a star is redder than expected we will
overestimate $A_V$).  This matrix, obviously independent of the star
considered, can easily be determined by using a control field, where
the extinction can be neglected, and measuring the scatters of star
colors.

The second matrix, $\Cov_{ij}(\epsilon)$, is related to photometric
errors, and thus changes for each star.  This matrix can be easily
calculated provided that an estimate of the photometric errors for
each star is available.  In the typical case of the three bands $J$,
$H$, and $K$ considered here, we can write
\begin{equation}
  \label{eq:10}
  \Cov_{ij}(\epsilon) = 
  \begin{pmatrix}
    \sigma^2_J + \sigma^2_H & - \sigma^2_H \\
    -\sigma^2_H & \sigma^2_H + \sigma^2_K
  \end{pmatrix} \; .
\end{equation}
Note that, while the photometric errors on different bands (namely,
$\sigma_J$, $\sigma_H$, and $\sigma_K$) can be taken to be
uncorrelated, errors on colors are not.

We can now minimize Eq.~\eqref{eq:9} with the requirement that
Eqs.~\eqref{eq:7} holds using Lagrange's multipliers.  This way we
reduce our problem to the solution $(b_1, b_2)$ of the linear system
\begin{equation}
  \label{eq:11}
  \begin{pmatrix}
    C_{11} & C_{12} & -k_1 \\
    C_{21} & C_{22} & -k_2 \\
    -k_1   & -k_2   & 0
  \end{pmatrix}
  \begin{pmatrix}
    b_1 \\ b_2 \\ \lambda
  \end{pmatrix}
  =
  \begin{pmatrix}
    0 \\ 0 \\ -1
  \end{pmatrix}
  \; ,
\end{equation}
where the matrix $C_{ij} = \Cov_{ij}(c^\mathrm{tr}) + \Cov_{ij}
(\epsilon)$ is the sum of the two covariances.  The parameter
$\lambda$ in the previous equation is Lagrange's multiplier and,
for our purposes, its value is uninteresting.  The solution of this
system is then, in matrix notation,
\begin{equation}
  \label{eq:12}
  \vec b = \frac{C^{-1} \cdot \vec k}{\vec k \cdot C^{-1} \cdot \vec
  k} \; ,
\end{equation}
where $\vec b = (b_1, b_2)$ and $\vec k = (k_1, k_2)$.  The final
expression for the optimal estimator is thus
\begin{equation}
  \label{eq:13}
  \hat A_V = b_1 \bigl[ c_1^\mathrm{obs} - \langle c_1^\mathrm{tr}
  \rangle \bigr] + b_2 \bigl[ c_2^\mathrm{obs} - \langle c_2^\mathrm{tr}
  \rangle \bigr] \; ,
\end{equation}
with $\vec b$ given by Eq.~\eqref{eq:12}.  Its variance is given by
Eq.~\eqref{eq:9}, or equivalently $\Var \bigl( \hat A_V \bigr) = \vec
b \cdot C \cdot \vec b$.

A nice feature of this technique is that it can be applied without
significant modifications in the case where one of the bands is not
observed.  Suppose, for example, that the $J$ band is not available.
Then we can assume an arbitrary value for this band and set the
corresponding error $\sigma_J$ to an extremely large value.  If we
``blindly'' apply Eqs.~\eqref{eq:12} and \eqref{eq:13}, we obtain a
matrix $C^{-1}$ with only a single non-vanishing element at the
position $(2,2)$.  In other words, the use of a large error on $J$
automatically suppresses the use of this band in the evaluation of
$\hat A_V$.  Interestingly, the same technique can be used if the
missing band is $H$, which contributes to both colors $c_1 = J - H$
and $c_2 = H - K$: In this case the estimator will be composed only of
the combination $c_1 + c_2 = J - K$, thus avoiding the use of the
missing $H$ band.

\subsection{Spatial smoothing}
\label{sec:spatial-smoothing}

So far we have considered only a single star.  In order to obtain a
smooth extinction map with high signal-to-noise ratio, \textsc{Nicer}
applies a \textit{spatial smoothing\/} to angularly close stars.  The
smoothing is performed using three different techniques, described
below in individual subsections.

The choice of the smoothing technique is important for two main
reasons: (i) The final map has a signal-to-noise ratio that strongly
depends on the smoothing used; (ii) The smoothing is responsible for
the selection of background stars.  As already pointed out,
\textsc{Nicer} tries to deal with both problems at the same time.

\subsubsection{Weighted mean}
\label{sec:weighted-mean}

The simplest way to obtain a smooth mass map from individual estimates
for $A_V$ is to use a weighted mean.  More precisely, we use the
values of $\hat A_V$ obtained for stars close to a direction $\vec
\theta$ on the sky to estimate the extinction on $\vec \theta$.
Calling $\vec\theta^{(n)}$ the position on the sky of the $n$-th star,
$\hat A_V^{(n)}$ its estimated extinction, and $\Var\bigl( \hat
A_V^{(n)} \bigr)$ the corresponding estimated variance, we write
\begin{equation}
  \label{eq:14}
  \hat A_V(\vec\theta) = \dfrac{\sum_{n=1}^N W^{(n)} \hat A_V^{(n)}
  }{\sum_{n=1}^N W^{(n)}} \; ,
\end{equation}
where $W^{(n)}$ is the weight for the $n$-th star, given by
\begin{equation}
  \label{eq:15}
  W^{(n)} = \frac{W\bigl( \vec\theta - \vec\theta^{(n)}
  \bigr)}{\Var\bigl( \hat A_V^{(n)} \bigr)} \; .
\end{equation}
Note that, in contrast to Eq.~\eqref{eq:3}, here the index $n$ runs
over \textit{all\/} $N$ observed stars in the field.  The weight
$W^{(n)}$ is the combination of two different factors: (i) A spatial
term, described by a \textit{weight function\/} $W$; (ii) An error
weight, proportional to $1 / \Var\bigl( \hat A_V \bigr)$.  The weight
function is usually chosen to be a Gaussian with appropriate width.
We stress that the choice of the characteristic length of the weight
function is a fundamental step in the reconstruction process, as shown
by a detailed statistical study of the statistical properties of the
smoothed map.  This has already been done in a different context
(\citealp{P2}; see also \citealp{P14}), and thus we just state the
main results obtained.  The size of the weight function directly
determines the \textit{correlation length\/} of the final extinction
map, or, equivalently, sets the scale of the smallest details we can
hope to observe on the map (every feature smaller than the
characteristic size of the weight function will be washed out).
Moreover, the larger the width of the weighting function, the larger
the \textit{effective\/} number of stars used for each point on the
map, and the smaller the error on the extinction.  In other words, we
can decide whether we want a high signal-to-noise ratio map with low
resolution, or a more noisy map with higher resolution.  A local
estimate for the error $\sigma_{\hat A_V}$ is given by
\begin{equation}
  \label{eq:16}
  \sigma_{\hat A_V}^2 (\vec\theta) = \frac{\sum_{n=1}^N \bigl( W^{(n)}
  \bigr)^2 \Var \bigl( A_V^{(n)} \bigr)}{\sum_{n=1}^N \bigl( W^{(n)}
  \bigr)^2} \; ,
\end{equation}
where $\Var \bigl( A_V^{(n)} \bigr)$ is the expected variance for the
$n$-th star [see note after Eq.~\eqref{eq:13}].  Finally, we note that
the choice \eqref{eq:15} for the coefficients of the stars maximizes
the signal-to-noise ratio.

So far we have assumed that all stars are background objects with
respect to the cloud.  In reality, some of them could be in the
foreground and could thus contaminate our extinction estimate.  If the
number of foreground stars is not too large, we can ignore them and
proceed \textit{as if\/} all stars were in the background.  Because of
the linearity of the estimator \eqref{eq:14}, we can correct for the
dilution of the signal \textit{a posteriori\/} by multiplying the map
$\hat A_V(\vec\theta)$ by the factor $N / N_\mathrm{back}$, where
$N_\mathrm{back}$ is the estimated number of background stars.
Actually, this simple technique is often not very efficient and can
also introduce biases for dense clouds (see below
Sect.~\ref{sec:discussion}).  For this reasons, it might be preferable
to use other smoothing techniques described below.

\subsubsection{Sigma-clipping}
\label{sec:sigma-clipping}

Foreground stars do not contribute to the signal, still contribute to
the noise.  Hence, it is desirable to exclude in the average
\eqref{eq:14} those stars that are suspected to be in the foreground
because of their low extinction.  This can be easily obtained by using
a sigma-clipped or by a median estimator (described below in
Sect.~\ref{sec:weighted-median}).

In order to implement sigma-clipping, we make a first estimate of
$A_V$ using Eq.~\eqref{eq:14} and calculate the expected variance
from Eq.~\eqref{eq:16}.  Then we repeat the entire process using only
those stars for which $\hat A^{(n)}_V$ does not differ by more than a
factor $3 \sigma_{\hat A_V}$ from the first estimate of $A_V$
evaluated at $\vec\theta^{(n)}$, i.e. $\hat A_V \bigl(
\vec\theta^{(n)} \bigr)$.  This procedure quickly converges to a pair
of values, the measured extinction $\hat A_V(\vec\theta)$ and the
related error, which we take as fiducial values.

In order to verify the goodness of our procedure, we can evaluate the
\textit{observed\/} variance on the data:
\begin{equation}
  \label{eq:17}
  \hat \sigma_{\hat A_V}^2 (\vec\theta) = \frac{\sum_{n=1}^N \bigl( W^{(n)}
  \bigr)^2 \bigl( \hat A_V^{(n)} - \hat A_V
  \bigr)^2}{\sum_{n=1}^N \bigl[ W^{(n)} \bigr]^2} \; .
\end{equation}
Note that this variance is defined in a different way with respect to
Eq.~\eqref{eq:16}.  Assuming that the photometric errors on the
various bands have been correctly evaluated, we expect $\hat
\sigma_{\hat A_V} \geq \sigma_{\hat A_V}$, the two variances being
equal only if all stars are subject to the same extinction, i.e.\ if
there are no foreground stars and the cloud is perfectly homogeneous.
Actually, the difference between $\hat \sigma_{\hat A_V}$ and
$\sigma_{\hat A_V}$ can be used to obtain precious information on the
substructure of the molecular cloud \citep[see,
e.g.][]{1998A&A...338..723J}.

\subsubsection{Weighted median}
\label{sec:weighted-median}

Alternatively, we can use a \textit{weighted median\/} in order to
make the extinction map.  This quantity is defined as the solution
$\hat A(\vec\theta)$ of the non-linear equation
\begin{equation}
  \label{eq:18}
  \sum_{n=0}^N W^{(n)} \sgn\bigl( \hat A(\vec\theta) - \hat A^{(n)}
  \bigr) = 0 \; ,
\end{equation}
where $\sgn$ is the sign function ($\sgn x = \{+1, 0, -1\}$ depending
on the sign of $x$).  Since in general the condition \eqref{eq:18}
cannot be satisfied exactly, a linear interpolation is used (see
\citealp{2000A&A...363..401L} for a similar example of application of
a weighted median).

The use of the median has several advantages with respect to the
simple mean.  Probably, the most interesting is the fact that the
median is a \textit{robust\/} estimator, since it is basically
unaffected by outliers.  This property is particularly useful in our
case, since it provides an efficient method to remove foreground
objects from our map.  On the other hand, the median has the
significant disadvantage of having noise properties difficult to
study.  We also note that the implementation of the median is not as
efficient as the mean, but this should be considered a minor problem.

\subsubsection{Discussion}
\label{sec:discussion}

In the previous subsections we have described three simple smoothing
techniques used by \textsc{Nicer}.  An important point to note is that
there is not really a single \textit{best method\/} to perform the
spatial smoothing.  In some conditions, several subtle problems, in
fact, may make each technique described above inappropriate, noisy, or
even biased.

As already noted by \citet{1997A&A...319..948T}, the simple or
weighted mean (as described in Sect.~\ref{sec:weighted-mean}) can be
severely biased in high-extinction regions.  In fact, because of
extinction, the density of background stars decreases as the cloud
column density increases, while the density of foreground stars is
left unchanged.  As a result, in the dense regions foreground stars
represent a larger fraction of the local number of stars with respect
to less dense regions.  This way, we could under-estimate the column
density in the central regions of clouds.  Such a bias can generally
be ignored for very nearby clouds.  In other cases, the bias might be
reduced using the median or the sigma-clipping methods.

On the other hand, substructure on the scale of the weighting function
can also be a problem.  In case of the simple weighted mean, the
smoothing procedure is straightforward to understand (at least at the
simple level kept in the discussion here; see however \citealp{P14}):
The final map obtained is expected to be the original, true extinction
of the cloud convolved with the weight function used.  In the case of
sigma-clipping or median smoothing, things are much more complicated.
Suppose, for example, that the cloud does not present any significant
substructure on the scale of the weighting function, except for small
``holes'' with very low absorption.  In this case, we would identify
stars observed through these holes as foreground objects, although
they are in reality in the background.  As a result, we would
completely miss the holes, and thus overestimate the cloud column
density.  Similarly, for a cloud with substructures consisting of
dense globules, we would underestimate the column density (note that
both sigma-clipping and median are ``symmetric,'' in the sense that
they discard outliers with both large or small $A_V$).

In summary, each smoothing technique has advantages and disadvantages
(this is the ultimate reason to include all of them in
\textsc{Nicer}).  A key role is played by the fraction of foreground
stars (and thus the distance of the cloud), and by the substructure of
the cloud.  It is in any case always advisable to use all smoothing
techniques and to check the consistency of the maps obtained.

\section{\textsc{Nicer} at work: The Orion region from 2MASS data}
\label{sec:nicer-at-work}

The Two Micron All Sky Survey (2MASS) offers a unique opportunity to
test our algorithm.  Conceived about a decade ago
\citep{1994ExA.....3...65K}, the 2MASS project is an all-sky-survey
obtained from two 1.3-m telescopes in the $J$ ($1.25\ \mathrm{\mu
  m}$), $H$ ($1.65 \ \mathrm{\mu m}$), and $K_\mathrm{s}$ band ($2.17
\ \mathrm{\mu m}$).  Although the surveys has been completed, only
about $47\%$ of the sky is at present publicly available in the Second
Incremental Release.\footnote{See
  \texttt{http://www.ipac.caltech.edu/2mass/}.}\@

We selected from the 2MASS archive a large area around the Orion and
MonR2 nebulae, with galactic coordinates
\begin{align}
  \label{eq:19}
  200^\circ <{} & l < 230^\circ \; , & -30^\circ <{} & b < -5^\circ \; .
\end{align}
The Orion region is a good candidate for our study for several
reasons.  The density of background stars, although not particularly
high, is enough to perform a good analysis.  Moreover, the 2MASS
second incremental release covers this region very well, with only a
small number of ``holes'' (except for the Nord-West part).  Finally,
there are already a number of studies in the literature which can be
used to test our results (e.g., \citealp{2000AJ....120.3139C} uses the
same data set to investigate the spatial distribution of young stars
in these clouds and presents averaged $J-K_s$ color maps of this
region).

\begin{figure}[t!]
 \begin{center}
   \begin{center}
     \ {}
     \vskip 8truecm
     \ {} 
   \end{center}
   \begin{center}
     \ {}
     \vskip 8truecm
     \ {} 
   \end{center}
   \caption{\textbf{Top:} Color-color diagram for stars in the control
     field, made from the colors of $6{,}000$ stars.  The ellipse
     represents a $3 \sigma$ confidence region.  \textbf{Bottom:}
     Color-color diagram for stars in the central region of the cloud.
     The reddening due to the dust in the cloud significantly
     stretches the locus of stars in the plot.}
   \label{fig:3}
 \end{center}
\end{figure}

For each point source in this coordinate range we retrieved the
magnitude on the three bands (fields \texttt{x\_mag}), the relative
total errors (\texttt{x\_msigcom}), and some additional flags used to
classify the goodness and the reliability of the detection
(\texttt{cc\_flg} and \texttt{rd\_flg}).  Possible artifacts (objects
with \texttt{cc\_flg}${} \neq 0$) were excluded from the analysis.  We
then applied our dedicated pipeline, as described in the following
sections.

\subsection{Preliminary analysis}
\label{sec:preliminary-analysis-1}

\begin{figure*}[t!]
 \begin{center}
   \includegraphics[angle=-90, width=\textwidth, keepaspectratio]{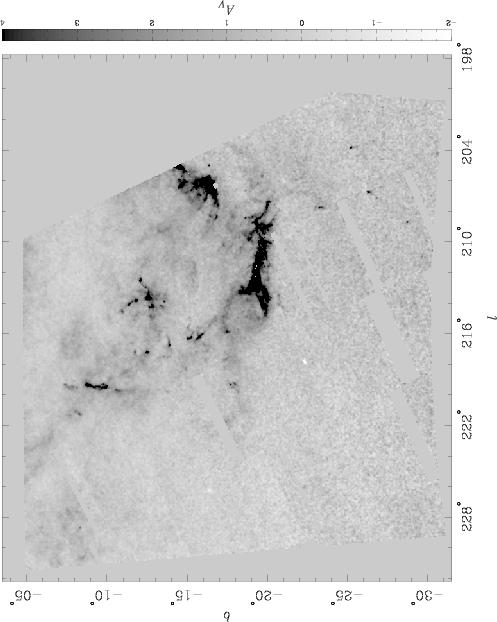}
   \caption{The total region mapped using \textsc{Nicer}.  The cuts
     used in $A_V$ emphasize the faint, diffuse halos around the main
     cores.  The low noise of this map allows us to detect a $A_V =
     0.5$ extinction with $3 \, \sigma$ confidence level.  The higher
     noise observed in the southern part is due to the smaller star
     density.}
   \label{fig:4}
 \end{center}
\end{figure*}

Given the area on the sky and the number of stars, the pipeline
``suggested'' to perform the analysis using a grid of $820 \times 700$
points, with scale $2.4'$ per pixel, and with Gaussian smoothing
characterized by $\mbox{FWHM} = 4.8'$.  By default, the pipeline uses
a $FWHM$ of 2 pixels, and about 15 ``effective'' stars per point
(i.e., about 15 stars contribute to the signal for each point).  We
then made a preliminary analysis using standard values for the average
colors and color scatters, thus obtaining a first extinction map for
the region.  The map obtained was used only as a first check for the
basic parameters chosen, to identify a control field, and to obtain
the photometric properties of stars in our sample.  In particular,
inside the boundaries \eqref{eq:9}, we found a region in the
South-East of the map which does not show any significant extinction
and which thus has been selected as control field (see
Fig.~\ref{fig:4}).  We then used the color of stars in this control
field to set the photometric parameters needed in our algorithm,
namely the average colors of stars (i.e., $\langle J - H\rangle$ and
$\langle H - K\rangle$) and the scatter in colors (or, more precisely,
the covariance matrix, which describes also correlations between
colors).  This step was performed in our pipeline by flagging stars
inside the control region (which can be defined using the standard
\texttt{SAOimage} format for region files) and by making a statistical
analysis on the photometric data of flagged stars.  Note that here
only the 2MASS photometric data were used, and not the preliminary map
(this map was used only to select the control field).  As a result, an
inaccurate choice for the initial constants of Eq.~\eqref{eq:6} to
make the preliminary map had no influence on the statistical
parameters obtained.

We observe that the use of a large survey such as 2MASS has the
significant advantage with respect to normal observations of letting
us study in detail the halos of molecular clouds.  As a result, the
choice for the control field is very robust, in the sense that even
regions with extremely low extinction can be avoided.  We also stress
that this is possible because of the availability of data on large
areas.  Fig.~\ref{fig:3} shows the color dispersion of stars in the
control field, together with the $3 \sigma$ elliptical confidence
region.

\subsection{Final map}
\label{sec:final-map-1}

\begin{figure*}[t!]
 \begin{center}
   \includegraphics[angle=-90, width=\textwidth, keepaspectratio]{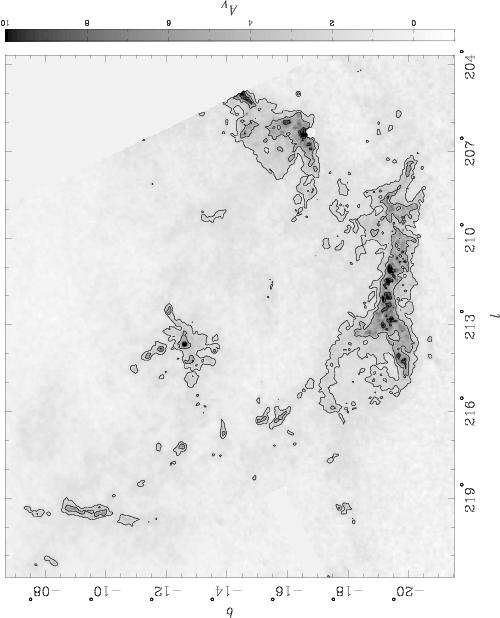}
   \caption{Zoom of the central region of the \textsc{Nicer}
     extinction map.  Cuts in this map emphasize the faint, diffuse
     halos around the main cores.  Contours are displayed at $A_V =
     \{2,4,8\}$ magnitudes.  Note that the maximum value obtained for
     $A_V$ is $17.7$ magnitudes.}
   \label{fig:5}
 \end{center}
\end{figure*}

\begin{figure*}[t!]
 \begin{center}
   \includegraphics[angle=-90, width=\textwidth, keepaspectratio]{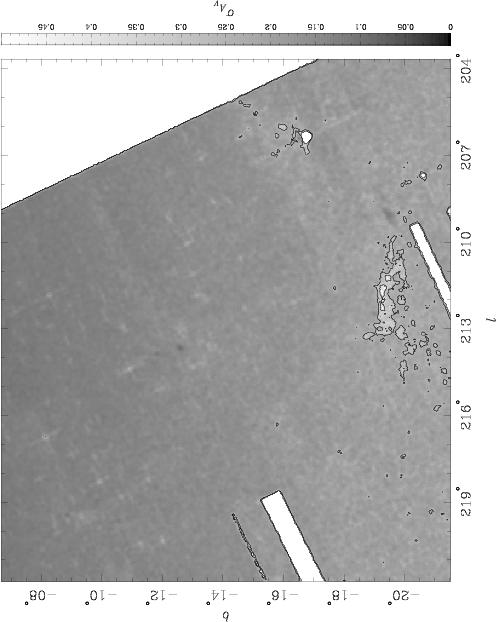}
   \caption{The noise estimate $\sigma_{\hat A_V}$ [cf.\
     Eq.~\eqref{eq:16}] for the central region shown in
     Fig.~\ref{fig:5}.  The average noise ranges from about $0.15$ to
     the top of the figure to about $0.25$ to the bottom (where the
     density of background stars is smaller because of the increase in
     $|b|$); contours are plotted at $\sigma_{\hat A_V} = \{ 0.25,
     0.35 \}$.  In the white regions the error diverges because of the
     lack of data.  Increase in noise is observed in high absorption
     regions (because of the decrease in density of stars), and around
     bright stars in the top of the figure (which have been masked by
     the 2MASS pipeline).  The white dot on the right is due to bright
     star zeta Ori (V $=$ 1.79 mag).  Except for these features, the
     overall noise level is rather uniform.}
   \label{fig:6}
 \end{center}
\end{figure*}

\begin{figure*}[t!]
 \begin{center}
   \includegraphics[angle=-90, width=\textwidth,
   keepaspectratio]{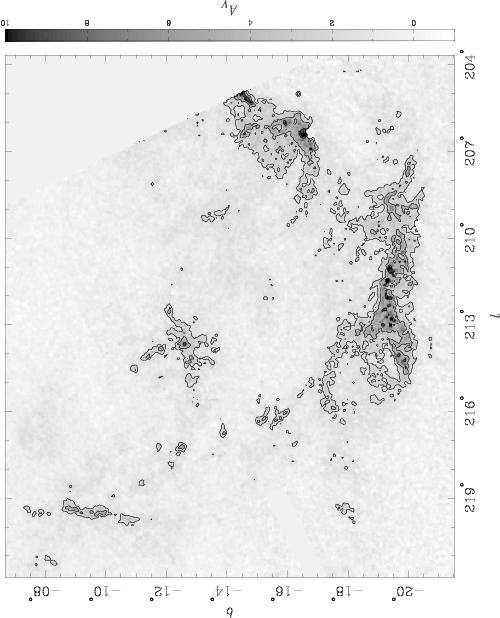}
   \caption{The same region shown in Fig.~\ref{fig:5} 
     but with the extinction map obtained using the standard method
     \citep{1994ApJ...429..694L}.  A quantitative analysis of the
     noise, carried out in regions with the lowest column density,
     shows that the variance of this map is about a factor two larger
     than the one shown in Fig.~\ref{fig:5}.  Note also that, although
     the contours levels are the same as in Fig.~\ref{fig:5}, they
     appear much less smooth.  This is another indication of the
     higher noise level of this map (rather than of intrinsic cloud
     structure; note that the resolution is the same for all images).}
   \label{fig:7}
 \end{center}
\end{figure*}

\begin{figure}[t!]
 \begin{center}
   \includegraphics[angle=-90, width=4truecm]{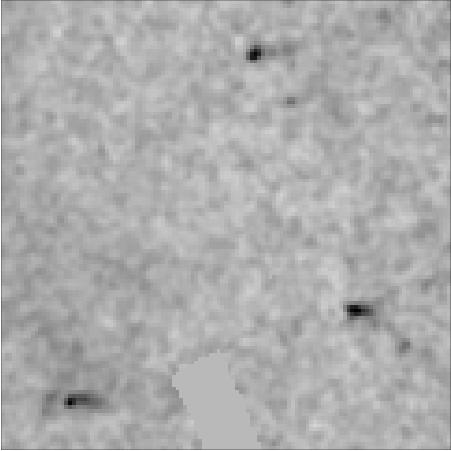}%
   \hspace{0.5truecm}%
   \includegraphics[angle=-90, width=4truecm]{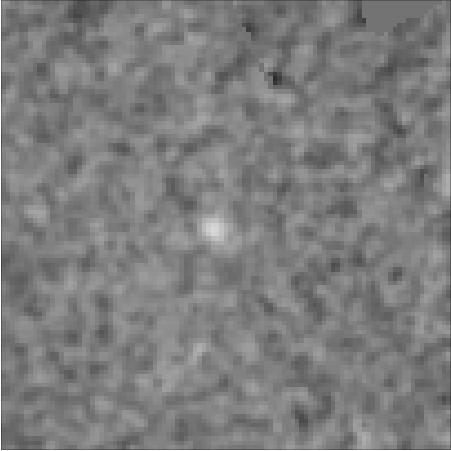}%
   \caption{\textbf{Left:} Peculiar cometary structures are observed
     south of the main cloud. The size of the image is $5\times 5
     \mbox{ deg}^2$.  \textbf{Right: } The only significant
     ``negative'' extinction corresponds to the open cluster NGC 2204,
     an open cluster in the halo (Hawarden 1976), whose average color
     is bluer than the average galactic field color.  The size of the
     image is $4 \times 4 \mbox{ deg}^2$.}
   \label{fig:8}
 \end{center}
\end{figure}

The final extinction map, shown in Fig.~\ref{fig:4}, was performed
using the method discussed in this paper.  Stars in the control field
was used to set the intrinsic colors and color dispersions of (i.e.,
to evaluate the quantities $\langle c_i \rangle$ and
$\Cov_{ij}(c^\mathrm{tr})$ described in
Sect.~\ref{sec:local-absorption}).  For each star in our region, we
evaluated the column density using Eqs.~\eqref{eq:12} and
\eqref{eq:13}.  Finally, column densities for close stars was averaged
out using three different schemes, namely simple (weighted) mean,
sigma clipping, and median.

Figures~\ref{fig:4} and \ref{fig:5} show the result obtained using a
Gaussian smoothing characterized by $\textmd{FVHM} = 5'$ with
$\sigma$-clipping (other smoothing schemes gives similar results, and
differences cannot generally be seen by eye).  Note that use of our
optimized method, combined with the quality of the 2MASS data, allows
us to detect at an unprecedented level of detail the faint, extended
halos on which the main cores are embedded.  At the smoothing size
used, we reach the excellent sensitivity $A_V = 0.13$ at $1 \, \sigma$
level.

Since the focus of this paper is mainly on the optimized
reconstruction method, we will defer a discussion of the results
obtained on this and on similar cloud complexes to a future paper.
However, we want to point out a couple of peculiar features of the map
obtained.  First, we detect a single region with significantly bluer
color (see Fig.~\ref{fig:8}, right).  This region corresponds to the
open cluster NGC~2204, whose blue stars contaminate our map thus
simulating a ``negative'' extinction.  Note also the cometary
structures observed south to the Orion cloud (Fig.~\ref{fig:8}, left).
In these condensations we observe extinction $A_V > 6$ on scales close
to our smoothing length.  Clearly, it would be worthwhile to study
these regions at higher resolution.

In order to better appreciate the advantages obtained using the
optimized technique described in this paper, we also carried out the
analysis on the same field using only the weighted average on the $H -
K$ color for stars with relatively low photometric errors (we required
errors on $H$ and $K$ to be smaller than $0.15$ magnitudes).
Figure~\ref{fig:7} shows the result obtained with this standard
method, and should be compared with Fig.~\ref{fig:5}.
\begin{itemize}
\item We first note that, apart from the different noise level, the
  two maps basically give the same values for the column density.
  This proves that the new method described in this paper is reliable
  and unbiased.
\item The simple $H - K$ map is significantly noisier than the one
  obtained using the optimized method.  Note that the resolution, set
  by the scale of the Gaussian smoothing, is the same in both maps.  A
  quantitative analysis shows that with the use of the optimized
  method we gain a factor of two on the variance of the map.  In other
  words, the new technique is able to reach the same signal-to-noise
  ratio of the standard method using only half of the stars.
\item Because of the reduced noise level, in Fig.~\ref{fig:4} we are
  able to distinguish regions with extremely low column density ($A_V
  \simeq 0.5$ magnitudes), which would be otherwise undetectable.
\end{itemize}
We note that the exact gain obtained with the optimized method
\textit{strongly\/} depends on the individual intrinsic scatter of
stars on each color.  This last point depends on several factors, such
as the depth of observations, the galactic latitude of the field, and
the colors used.  On the other hand, given the excellent results
obtained with the 2MASS archive, we are confident that \textsc{Nicer}
can produce extinction maps with significantly higher signal-to-noise
ratio for any infrared data.  Finally, we stress that although the
method has been presented here for the $J$, $H$, and $K$ bands, it can
actually be applied to any set of bands.

\section{Conclusions}
\label{sec:conclusions}

In this paper we have presented an optimized technique to produce
highly accurate extinction maps from multi-band near-infrared
photometric data.  The method, which is a natural generalization of
the near infrared color excess method of \citet{1994ApJ...429..694L},
is able to produce significantly less noisy (and thus more accurate)
extinction maps taking advantage of all bands available.  A first
example of application of this new technique to 2MASS data has shown
an improvement with respect to the standard \textsc{Nice} algorithm of
a factor $2$ on the noise variance.  This way, we have been able to
detect extended diffuse halos down to $A_V \simeq 0.5$ magnitudes.

\acknowledgements We thank Charles Lada and Lindsay King for helpful
discussions and observations on this work.  We also thank the referee,
Doug Johnstone, for useful suggestions that improved the paper.  This
publication makes use of data products from the Two Micron All Sky
Survey, which is a joint project of the University of Massachusetts
and the Infrared Processing and Analysis Center, funded by the
National Aeronautics and Space Administration and the National Science
Foundation.

\bibliographystyle{apj}
\newcommand{\apj}[0]{ApJ}
\newcommand{\apjl}[0]{ApJ}
\newcommand{\aj}[0]{AJ}
\newcommand{\aap}[0]{A\&A}
\newcommand{\aaps}[0]{A\&ASS}
\newcommand{\apss}[0]{ApSS}
\newcommand{\apjs}[0]{ApJSS}
\newcommand{\mnras}[0]{MNRAS}
\newcommand{\pasp}[0]{PASP}
\newcommand{\pasj}[0]{PASJ}
\newcommand{\nat}[0]{Nature}
\newcommand{\araa}[0]{ARAA}
\bibliography{refs.bib}

\end{document}